\documentclass[pra,twocolumn,reprint]{revtex4-1}
\usepackage{amsmath,amssymb,amsthm,bm,braket,ascmac,bbm,color}
\usepackage{graphicx}

\theoremstyle{definition}

\newcommand{\dd}{\mathrm{d}}
\newcommand{\ee}{\mathrm{e}}
\newcommand{\ii}{\mathrm{i}}

\newcommand{\nlat}{p}
\newcommand{\resol}{\mathcal{K}(t)}

\newcommand{\FG}{\mathrm{FG}}

\newcommand{\besdiff}{J}
\newcommand{\current}{\mathcal{J}}

\begin{document}
\title{
Revisiting the Floquet-Bloch theory
for an exactly solvable model of one-dimensional crystals
in strong laser fields
}

\author{Tatsuhiko N. Ikeda}
\affiliation{Institute for Solid State Physics, University of Tokyo, Kashiwa, Chiba 277-8581, Japan}
\date{\today}
\begin{abstract}
We revisit the Floquet-Bloch eigenstates of a one-dimensional electron gas
in the presence of the periodic Kronig-Penney potential
and an oscillating electric field.
Considering the appropriate boundary conditions for the wave function
and its derivative,
we derive the determining equations for the Floquet-Bloch eigenstates,
which are represented by a single-infinite matrix
rather than a double-infinite matrix needed for a generic potential.
We numerically solve these equations, showing that
there appear anticrossings at the crossing points of the different Floquet bands
as well as the band gaps at the edges and the center of the Brillouin zone.
We also calculate the high-harmonic components of the electric current carried by the
Floquet-Bloch eigenstates, showing that
the harmonic spectrum shows a plateau for a strong electric field.
\end{abstract}
\maketitle

\section{Introduction}
High-harmonic generation (HHG) in bulk crystals
has recently attracted much attention
owing to its successful observation
in the strong laser field~\cite{Ghimire2011,Schubert2014,Hohenleutner2015,Luu2015,Ndabashimiye2016,Ghimire2014}.
HHG in solids in the nonperturbative regime thus achieved
has shown different characters from that in atomic gases~\cite{McPherson1987,Ferray1988},
such as the multiple plateaus~\cite{Ndabashimiye2016} and the linear scaling of the high-energy cutoff
with the electric field rather than the laser intensity~\cite{Ghimire2011,Schubert2014,Luu2015}.
Although the microscopic mechanism of HHG in solids is still under an active debate~\cite{Korbman2013,Higuchi2014,Vampa2014,Vampa2015},
it is pointed out that
the interband transitions between multiple bands
play an essential role~\cite{Golde2008,Ndabashimiye2016,McDonald2015,Hawkins2015,Wu2015,Du2017,Wu2016,Ikemachi2017}

The most fundamental model to discuss HHG in solids
is  a one-dimensional electron gas in the presence of both
a periodic lattice potential and an oscillating electric field.
Recently, this model has been shown to reproduce some aspects of
experimental results by numerical simulations, in which
the time-dependent Schr\"{o}dinger equation is explicitly solved at each time step~\cite{Wu2015,Du2017,Ikemachi2017,Jia2017,Ikemachi2017a}.
A complementary approach to this problem is the Floquet-Bloch theory~\cite{Tzoar1975,Faisal1989,Faisal1991,Faisal1997,Gupta2003,Alon2004,Faisal2005},
which utilizes the periodicity both in time and space to obtain the eigenstates
for the time-dependent Schr\"{o}dinger equation.
In this approach, one needs to treat a double-infinite Hamiltonian matrix,
where its column or row is labeled by two integers corresponding to
the high harmonics for the time and space oscillation of the wave function.
Although approximate solutions are obtained numerically~\cite{Tzoar1975},
the numerical cost grows rapidly to achieve high accuracy.

The Kronig-Penney, or a square wave, potential~\cite{Kronig1931} simplifies the problem significantly.
This potential has traditionally deepened our understanding of the band gaps in solids
because it is analytically solvable in the limit of each square approaching a delta function (see Eq.~\eqref{eq:kppot}).
The analytical approach to the Floquet-Bloch theory for this potential
was discussed by Faisal and Genieser~\cite{Faisal1989,Faisal1991},
who proposed that the problem of the double-infinite Hamiltonian matrix can be reduced
to that of a single-infinite matrix and the calculation cost is thereby greatly decreased.
However, their result contains mistakes 
stemming from their misunderstanding that
the Floquet Green function is obtained by a simple generalization
of the time-independent problem.
Also, the high-harmonic components of the electric current
have not been obtained in the delta-function limit.

In this paper,
we revisit the Floquet-Bloch theory for the one-dimensional electron gas
with the Kronig-Penney potential in the delta-function limit
and analytically derive the correct determining equation for the Floquet-Bloch eigenstates.
As suggested in Refs.~\cite{Faisal1989,Faisal1991},
this equation consists of a single-infinite matrix rather than a double-infinite one.
We then numerically solve the equation, obtaining the quasienergy dispersion.
In addition to the energy gaps
at the edges and the center of the Brillouin zone
that are already present in the oscillating electric field,
there appear anticrossings between different Floquet bands
owing to the interplay of the periodic potential and the oscillating electric field.
We then calculate the high-harmonic components of the electric current
carried by the Floquet-Bloch eigenstates that are obtained in this formalism.
We show that, for a strong laser field, the high-harmonic components
do not decay exponentially, but show a plateau.

The rest of this paper is organized as follows.
In Sec.~\ref{sec:formulation},
we formulate the problem that is addressed in this paper.
By invoking the Floquet theorem,
we derive an eigenvalue problem for a Floquet Hamiltonian.
In Sec.~\ref{sec:solution},
we invoke the Bloch theorem and derive the determining equation
for the quasienergy dispersion from the conditions for the connections
of the wave function and its derivative.
In Sec.~\ref{sec:props},
we solve the equation to obtain the quasienergy dispersion
and discuss the band gaps and the anticrossings between the Floquet bands.
We then calculate the high-harmonic components of the electric current for the Floquet-Bloch eigenstates thus obtained.
The conclusions of this paper are summarized and
some future perspectives are shown in Sec.~\ref{sec:conclusions}.
In Appendix,
we point out why the original derivation~\cite{Faisal1989,Faisal1991}
is not correct
while it works in the absence of the oscillating electric field.

\section{Formulation of the problem}\label{sec:formulation}
Let us begin by considering the following Hamiltonian
in the velocity gauge
\begin{align}\label{eq:Ham0}
\mathcal{H}(t) = \frac{1}{2}\left[-\ii \partial_x - eA(t)\right]^2 + V(x).
\end{align}
Here $e$ $(<0)$ denotes the charge of an electron,
and we work in the units of $\hbar=m=1$
and use the abbreviation $\partial_x = \partial/\partial x$ throughout this paper.
The vector potential 
\begin{align}
A(t) = A_0 \cos(\Omega t)
\end{align}
represents the uniform laser electric field along the $x$ axis,
and the periodic lattice potential $V(x)$
is taken to be the Kronig-Penney one
in the delta-function limit:
\begin{align}\label{eq:kppot}
V(x) = \frac{P}{2a}\sum_{\nlat =-\infty}^\infty \delta(x-\nlat a),
\end{align}
where $a$ and $P$ denote
the lattice constant
and the strength of the lattice potential, respectively.
We note that $A(t)$ and $V(x)$ are both periodic:
$A(t+T) = A(t)$ and $V(x+a) = V(x)$ with $T\equiv 2\pi /\Omega$.

Our aim is to solve
the time-dependent Schr\"{o}dinger equation for 
the Hamiltonian~\eqref{eq:Ham0}
which is equivalent to find the eigenstates with the zero eigenvalue for the operator
\begin{align}
\mathcal{K}_0(t) = \ii \partial_t - \mathcal{H}(t).
\end{align}
To this end,
we eliminate the term proportional to $A(t)^2$ from $\mathcal{H}(t)$
by making the phase transformation
\begin{align}
\mathcal{K}(t) &= \ee^{\ii \theta(t)} \mathcal{K}_0(t) \ee^{-\ii \theta(t)} = \ii \partial_t - H(t),\\
H(t) &=  -\frac{1}{2}\partial_x^2 +\ii  eA(t)\partial_x + V(x),\label{eq:ham_org}
\end{align}
where $\theta(t) \equiv  \exp( -\frac{\ii}{2}\int_0^t  A(s)^2 \dd s )$ and $\partial_t =\partial/\partial t$.
We note that the operator $\mathcal{K}(t)$ has discrete translation symmetries
in the $t$ and $x$ directions due to the periodicities of $A(t)$ and $V(x)$, respectively.

The discrete translation symmetry in time
simplifies the problem of finding zero modes of $\resol$.
This symmetry 
tells us that
an eigenstate $\Psi(x,t)$ of $\mathcal{K}(t)$
is written as $\Psi(x,t)=\ee^{-\ii E t} \Psi_E(x,t)$,
where $\Psi_E(x,t)$ is a periodic function $\Psi_E(x,t+T)=\Psi_E(x,t)$.
We refer to $E$ as the quasienergy
since $E$ can be taken in a certain region such as $[0,\Omega)$.
However,
we do not restrict $E$ on such a region in this paper,
but work in the extended zone scheme,
in which we have physically equivalent states that have equal $E$ modulo $\Omega$.
Expanding $\Psi_E(x,t)$ in the Fourier series,
we obtain the following expression for an eigenstate
\begin{align}\label{eq:floquetexp}
\Psi(x,t) = \ee^{-\ii Et} \sum_{n\in \mathbb{Z}} \psi_n(x)\ee^{\ii n \Omega t}.
\end{align}
Now the equation $\resol \Psi(x,t)=0$, which we aim to solve,
reduces to the following form:
\begin{align}
& \sum_{n}H^0_{mn}\psi_{n}(x) + V(x)\psi_m(x)= E\psi_m(x),\label{eq:schroe1}\\
&H^0_{mn}=\left(-\frac{1}{2}\partial_x^2+n\Omega\right)\delta_{mn} +\ii \Omega \frac{\alpha_0}{2} (\delta_{m,n+1}+\delta_{m,n-1})\partial_x,\label{eq:schroe2}
\end{align}
where $\alpha_0 \equiv eA_0/\Omega$
represents the coupling strength between the electron
and the oscillating electric field.

The discrete translation symmetry in space
also simplifies the problem,
but this is not enough for the complete solution in general.
A parallel argument on space reduces the continuous variable $x$ to an integer
and the remaining problem is to diagonalize a double-infinite matrix where
its row or column is characterized by a pair of integers
corresponding to the high harmonics for the time and space oscillation of the wave function.
Although we can numerically obtain approximate solutions for the problem,
the computational complexity grows rapidly in increasing the precision.

\section{Solution to the eigenvalue problem}\label{sec:solution}
The Kronig-Penney potential~\eqref{eq:kppot}
enables us to proceed further analytically.
In this section, we analyze the eigenvalue problem~\eqref{eq:schroe1} and \eqref{eq:schroe2}
in the real space,
and show that the quasienergy $E$ is obtained as a root of a secular equation
for a single-infinite matrix (see Eq.~\eqref{eq:secular} or \eqref{eq:secular2} below).

Before solving our problem, we consider the solution in the absence of the periodic potential:
\begin{align}\label{eq:eq0}
H^0\bm{\psi}(x) = E\bm{\psi}(x).
\end{align}
In this case, the normalizable eigenstate characterized by a real momentum $k$
and an integer $N$
is given by
\begin{align}\label{eq:sol0}
\varphi^{N,k}_n(x) = J_{n-N}(\alpha_0 k) \ee^{\ii k x}
\end{align}
and its eigenvalue of $H^0$ is
\begin{align}\label{eq:e0}
\epsilon_0^N(k)=\frac{k^2}{2}+N\Omega.
\end{align}
Here $J_n(x)$ denotes the Bessel function of the first kind.
One can easily confirm that Eq.~\eqref{eq:sol0} satisfies Eq.~\eqref{eq:eq0} by acting
$H^0$ onto $\varphi^{N,k}_n(x)$
and using the identity $J_{n+1}(x)+J_{n-1}(x)=2nJ_n(x)/x$~\cite{Abramowitz}.
We note that the eigenstates~\eqref{eq:sol0} are mutually orthogonal
and satisfy the completeness relation
\begin{align}
\sum_{N\in\mathbb{Z}}\int_{-\infty}^\infty \frac{\dd k}{2\pi} \varphi^{Nk}_n(x) \varphi^{Nk}_{n'}(x')^* =\delta_{nn'}\delta(x-x').
\end{align}
This can be proved by using Neumann's identities~\cite{Abramowitz}
\begin{align}\label{eq:neumann}
\sum_{m\in \mathbb{Z}} J_{n\pm m}(z) J_{m}(w) = J_{n}(z\mp w)
\end{align}
and $J_{n-m}(0)=\delta_{nm}$.

We treat the effects of the periodic potential $V(x)$
as appropriate boundary conditions for the unit cell.
the Bloch theorem tells us that an eigenstate $\bm{\psi}(x)$ satisfies
\begin{align}\label{eq:bc1}
\bm{\psi}(a) = \ee^{\ii K a} \bm{\psi}(0)
\end{align}
for a lattice momentum $K$ with $-\pi/a\le K < \pi/a$.
In addition, the Kronig-Penney potential~\eqref{eq:kppot} imposes
boundary conditions for the first derivatives.
By integrating both sides of Eq.~\eqref{eq:schroe1} over $x\in [-\epsilon,\epsilon]$
and taking the limit of $\epsilon\downarrow 0$,
we obtain
\begin{align}\label{eq:bc2}
\partial_x \bm{\psi}(0+) - \ee^{-\ii K a}\partial_x \bm{\psi}(a-)-\frac{P}{a}\bm{\psi}(0)=0,
\end{align}
where 
we have used
$\partial_x \bm{\psi}(0-)=\ee^{-\ii K a}\partial_x \bm{\psi}(a-)$
which the Bloch theorem implies.
Since $V(x)$ vanishes on $0<x<a$,
our problem is to solve
\begin{align}\label{eq:eqK}
H^0\bm{\psi}(x) = E(K)\bm{\psi}(x) \quad (0<x<a)
\end{align}
with the boundary conditions~\eqref{eq:bc1} and \eqref{eq:bc2}.
The notation $E(K)$ emphasizes the fact that
the quasienergy bands over the first Brillouin zone are obtained
by varying $K$ appearing in the boundary conditions~\eqref{eq:bc1} and \eqref{eq:bc2}.

The most general solution of Eq.~\eqref{eq:eqK} is of the form
\begin{align}\label{eq:generalsol}
\psi_n(x) = \sum_{N\in\mathbb{Z}}\left[ A_N \varphi_n^{N,k_N}(x) +B_N \varphi_n^{N,-k_N}(x) \right],
\end{align}
with
\begin{align}\label{eq:defkN}
k_N \equiv \sqrt{2[E(K)-N\Omega]}.
\end{align}
We note that
there is no problem if 
$k_N$ becomes imaginary since we do not now work on
$-\infty<x<\infty$, but a finite range $0<x<a$.

The coefficients $\{A_N\}_N$ and $\{B_N\}_N$ are determined by
the boundary conditions~\eqref{eq:bc1} and \eqref{eq:bc2}.
By substituting the general solution~\eqref{eq:generalsol} into these conditions,
we obtain the following homogeneous matrix equation
\begin{align}\label{eq:homoeq}
\sum_{N\in \mathbb{Z}}
\begin{pmatrix}
\mathcal{M}_{nN}^{11} & \mathcal{M}_{nN}^{12}\\
\mathcal{M}_{nN}^{21} & \mathcal{M}_{nN}^{22}
\end{pmatrix}
\begin{pmatrix} A_{N} \\ B_{N} \end{pmatrix}
=0,
\end{align}
where the infinite matrix $\mathcal{M}^{ij}_{nN}$
in a $2\times2$ block form
is defined as
\begin{align}\label{eq:MmatDef}
\begin{pmatrix}
\mathcal{M}_{nN}^{11} & \mathcal{M}_{nN}^{12}\\
\mathcal{M}_{nN}^{21} & \mathcal{M}_{nN}^{22}
\end{pmatrix}
=C_N 
\begin{pmatrix}
J_{n-N}(\alpha_0k_N) & 0 \\ 0 & J_{n-N}(-\alpha_0k_N)
\end{pmatrix}
\end{align}
with
\begin{align}
C_N=
\begin{pmatrix}
\ee^{\ii k_{N}a}-\ee^{\ii Ka} & \ee^{-\ii k_{N}a}-\ee^{\ii Ka} \\
\ii k_{N} (1-\ee^{\ii(k_{N}-K)a})-\frac{ P}{a} & -\ii k_{N}(1-\ee^{\ii (-k_{N}-K)a} ) -\frac{P}{a} 
\end{pmatrix}
\end{align}
Equation~\eqref{eq:homoeq} has a nontrivial solution when
$E(K)$ is chosen so that
\begin{align}\label{eq:secular}
\det \mathcal{M} =0,
\end{align}
and the coefficients $\{A_N\}_N$ and $\{B_N\}_N$
are determined up to the overall factor.
We note that, if $E(K)=M\Omega$ for an integer $M$ and thus $k_M=0$,
the $N=M$ components are considered to be eliminated from the matrix
since $\varphi_{n}^{M,0}$ cannot satisfy the boundary conditions for $P\neq0$.

When solved for $E(K)$ for each $K$,
the secular equation~\eqref{eq:secular}
gives the quasienergy dispersions
in the presence of the oscillating electric field.
The multiple solutions obtained for a given $K$
correspond to the different bands.
We note that the row and the column of $\mathcal{M}$
are labeled by the pair of $(n,i)$ and $(N,i)\in \mathbb{Z}\times \{1,2\}$,
which have been remarkably simplified.
As noted in the previous section,
if we had not used the explicit form of the Kronig-Penney potential~\eqref{eq:kppot}
but had only invoked the periodicity in space,
we would have obtained a matrix whose column or row
runs over $\mathbb{Z}\times\mathbb{Z}$ corresponding to
the high harmonics for the wave function oscillations in time and space.

Another representation 
for the secular equation~\eqref{eq:secular} is obtained,
where the column and the row of the matrix
is treated on equal footing.
We multiply both sides of Eq.~\eqref{eq:homoeq} by $J_{n-N'}(\alpha_0|k_{N'}|)$
and sum them over $n$,
obtaining
\begin{align}\label{eq:homoeq}
\sum_{N\in \mathbb{Z}}
\begin{pmatrix}
\mathcal{W}_{N'N}^{11} & \mathcal{W}_{N'N}^{12}\\
\mathcal{W}_{N'N}^{21} & \mathcal{W}_{N'N}^{22}
\end{pmatrix}
\begin{pmatrix} A_{N} \\ B_{N} \end{pmatrix}
=0,
\end{align}
where the matrix elements of $\mathcal{W}_{N'N}$
are defined by 
\begin{align}\label{eq:defWmat}
\begin{pmatrix}
\mathcal{W}_{N'N}^{11} & \mathcal{W}_{N'N}^{12}\\
\mathcal{W}_{N'N}^{21} & \mathcal{W}_{N'N}^{22}
\end{pmatrix}
=C_N 
\begin{pmatrix}
\besdiff_{N'N}^{(-)} & 0 \\ 0 & \besdiff_{N'N}^{(+)}
\end{pmatrix}
\end{align}
with $\besdiff^{(\pm)}_{N'N} \equiv J_{N-N'}(\alpha_0(|k_{N'}|\pm k_N))$.
Making use of $|k_{N'}|$ in the procedure is technically advantageous
since it ensures that $\det \mathcal{W}$ is real or pure imaginary for any value of $E(K)$.
We assume that the matrix $J_{n-N}(\alpha_0|k_{N}|)$
has a nonzero determinant
when $n$ and $N$ are regarded as the labels for its row and column, respectively.
Then the secular equation~\eqref{eq:secular} is equivalent to
\begin{align}\label{eq:secular2}
\det \mathcal{W}=0.
\end{align}

In the absence of the electric field,
the secular equations~\eqref{eq:secular} and \eqref{eq:secular2} reproduce
the well-known Kronig-Penney dispersion relation~\cite{Kronig1931}
\begin{align}\label{eq:kpdisp}
\cos Ka = \cos k_0 a +\frac{P}{2k_0 a} \sin k_0 a
\end{align}
with $k_0 = \sqrt{2E(K)}$.
To confirm this,
let us note that
the absence of the electric field implies that 
$N$ only takes 0 and
the Bessel functions $J_{n-N}(\pm\alpha_0)$ and $\besdiff^{(\pm)}_{NN'}$ are replaced by unity.
Then the secular equations read $\det C_0 = 0$,
which readily leads to Eq.~\eqref{eq:kpdisp}.

\section{Properties of the Floquet-Bloch eigenstates}\label{sec:props}

In this section, we numerically solve the equations derived in the previous section
and discuss the properties of the Floquet-Bloch eigenstates.
For simplicity, we set the lattice constant $a$ as unity
throughout this section.

\subsection{Absence of the periodic potential}
Before discussing the effects of the periodic potential,
we investigate the properties of the Floquet eigenstates~\eqref{eq:sol0}
in the absence of the periodic potential.

\begin{center}
\begin{figure}
\includegraphics[width=8.5cm]{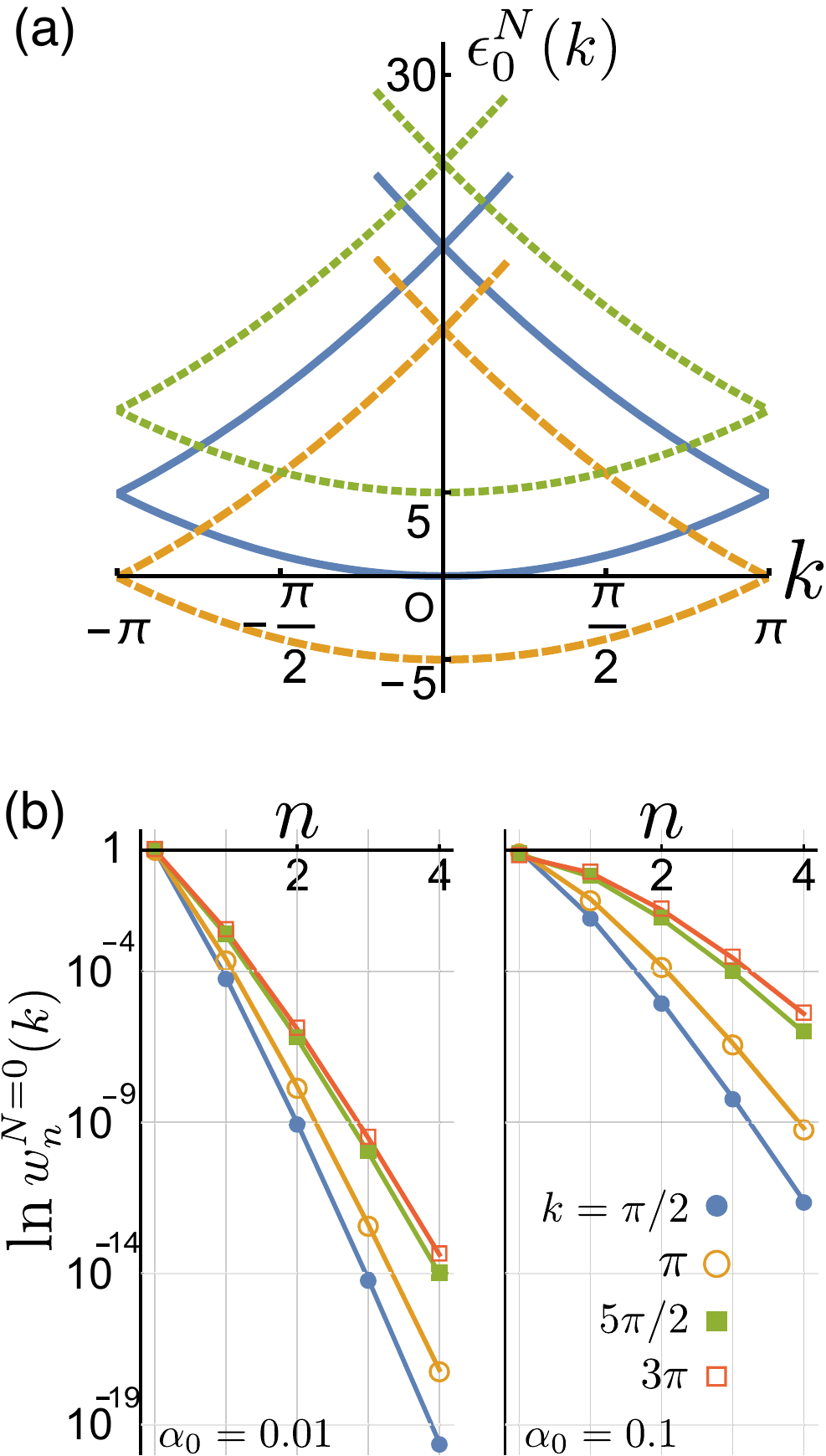}
\caption{
(Color Online)
(a) The quasienergy $\epsilon_0^N(k)$ [Eq.~\eqref{eq:e0}]
for $N=0$ (solid), $-1$ (dashed), and $1$ (dotted),
in the absence of the periodic potential, $P=0$,
and $\Omega=5$.
The horizontal axis represents the momentum $k$ modulo $2\pi$.
(b) The high-harmonic distribution $w^0_n(k)$ [Eq.~\eqref{eq:dist0def}] for $\alpha_0=0.01$ (left) and 0.1 (right).
Each dataset corresponds to
the momentum
$k=\pi/2$ (filled circle), $\pi$ (open circle),
$5\pi/2$ (filled square), and $3\pi$ (open square).
}
\label{fig:fig1}
\end{figure}
\end{center}

The quasienergy of $\varphi^{N,k}$ is given by 
Eq.~\eqref{eq:e0},
which is illustrated in Fig.~\ref{fig:fig1} (a).
The horizontal axis
of Fig.~\ref{fig:fig1} (a) shows $k$ modulo $2\pi$
for the comparison below to $E(K)$
in the presence of the periodic potential.
The quasienergy with $N=0$ is identical to $k^2/2$
in spite of the coupling to the oscillating electric field.
In addition to the dispersion relation $k^2/2$ for $N=0$,
we have an infinite number of replicas with equal spacings $\Omega$
since we work in the extended zone scheme as noted above.
In the following, we refer to the set of (quasi)energy for a given $N$
as the Floquet band.

Now we define
the high-harmonic distribution:
\begin{align}\label{eq:dist0def}
w_n^N(k) \equiv |\varphi^{N,k}_n(x)|^2 = J_{n-N}(\alpha_0k)^2.
\end{align}
This represents the weight of the Floquet eigenstate $\varphi^{N,k}_n(x)$
on the oscillating component with frequency $\epsilon^N_0(k) - n\Omega$
(see Eq.~\eqref{eq:floquetexp}).
We note that $w_n(k)$ is normalized as
\begin{align}
\sum_{n\in\mathbb{Z}} w_n^N(k) = 1,
\end{align}
which follows from Eq.~\eqref{eq:neumann}.
Since $w_{n}^{N+1}(k)=w_{n-1}^N(k)$ holds true,
it is enough to discuss $w_{n}^{0}(k)$.
In addition,
we have $w_{-n}^{0}(k)=w_{n}^{0}(-k)=w_{n}^{0}(k)$,
and, hence,
the nontrivial information
is contained in $w_n^0(k)$ for, say, $n\ge0$ and $k\ge0$.
In contrast to the quasienergy,
the high-harmonic distribution depends on the coupling strength $\alpha_0$
between the electron and the electric field.

The high-harmonic distribution $w^0_n(k)$ is
illustrated in Fig.~\ref{fig:fig1} (b)
for several values of $k$.
First, we note
$w_n^0(k=0)=\delta_{n0}$, which follows from the definition~\eqref{eq:dist0def}.
This is because
the electric field couples to the electron through its momentum
and, thus, high-harmonic oscillation is not induced for the $k=0$ state.
Second,
as $k$ increases with $\alpha_0$ held fixed
or $\alpha_0$ does with $k$ held fixed,  
the width of the high-harmonic distribution $w_n^0(k)$ becomes larger.
This tendency is qualitatively consistent with
the fact the the coupling to the electric field is proportional to the momentum.
We will show below in Sec.~\ref{sec:hhc} that
the width of $w_n^0(k)$ is related to the high-harmonic components
of the electric current in the presence of the periodic potential.
We note that, as anticipated from the Bessel function in Eq.~\eqref{eq:dist0def},
$w_n^0(k)$ does not show a monotonous but an oscillatory behavior
for even larger values of $\alpha_0$ or $k$.

\subsection{Effects of the periodic potential}
The Floquet-band theorem~\cite{Faisal1989} holds true
also in the presence of the periodic potential.
Namely,
if $E(K)$ satisfies Eq.~\eqref{eq:secular2},
then $E(K)+M\Omega$ does as well for any integer $M$.
To prove this, let us suppose that Eq.~\eqref{eq:secular2} is satisfied
for $E(K)$
and ask if $\det \widetilde{\mathcal{W}}=0$,
where $\widetilde{\mathcal{M}}$ is defined by replacing
$k_N$ in $\mathcal{W}$ by $\tilde{k}_N=\sqrt{2[E(K)+M\Omega-N\Omega]}=k_{N-M}$.
In this replacement,
$\besdiff^{(\pm)}_{NN'}$
in $\mathcal{W}$ is replaced by $\widetilde{\besdiff}_{NN'}^{(\pm)}=J_{N'-N}(\alpha_0(|\tilde{k}_N|\pm\tilde{k}_{N'}))
=J_{(N'-M)-(N-M)}(\alpha_0(|k_{N-M}|-k_{N'-M}))=\besdiff^{(\pm)}_{N-M,N'-M}$.
Thus $\widetilde{\mathcal{W}}$ is obtained by shifting the labels of both the row and the column of $\mathcal{W}$ by $M$.
Since this shift does not change the value of the determinant,
we have obtained $\det \widetilde{\mathcal{W}}=0$.

Both in absence and presence of 
the coupling between the electron and the electric field,
the quasienergy dispersion $E(K)$ is symmetric about $K=0$.
To prove this, let us suppose that Eq.~\eqref{eq:secular} is satisfied for $E(K)$
and show that $E(K)$ also satisfies Eq.~\eqref{eq:secular}
with $K$ replaced by $-K$.
For this purpose, we try to transform the matrix $\mathcal{M}$ for $K$
into that for $-K$ by means of elementary row and column operations.
The concrete procedure is the following.
First we multiply every row in the upper (lower) blocks by $\ee^{-\ii Ka}$ $(\ee^{\ii Ka})$
and the $N$-th column in the left (right) blocks by $\ee^{-\ii k_N a}$ $(\ee^{\ii k_N a})$.
Second we add the $n$-th row of the upper left (right) block
of the resulting matrix multiplied by  $-\ee^{-\ii Ka}P/a$ $(-\ee^{\ii Ka}P/a)$
to the $n$-th rows of the lower left (right) blocks.
Third we interchange the $N$-th columns of the left and right blocks for each $N$.
In these three steps, the determinant is invariant or changes its sign depending on whether
$N$ is thought to be even or odd.
The matrix thus obtained differs from $\mathcal{M}$ for $-K$
only in that $J_{n-N}(\pm\alpha_0 k_N)$ appears in the opposite way in the form of Eq.~\eqref{eq:MmatDef}.
This difference does not matter when their determinants are compared, for $J_{n-N}(\pm\alpha_0 k_N)=(-1)^{n-N}J_{n-N}(\mp\alpha_0k_N)$.
Thus we have shown that, $E(K)$ satisfies $\det \mathcal{M}=0$ for $K$,
it also does for $-K$.

\begin{center}
\begin{figure}
\includegraphics[width=8.5cm]{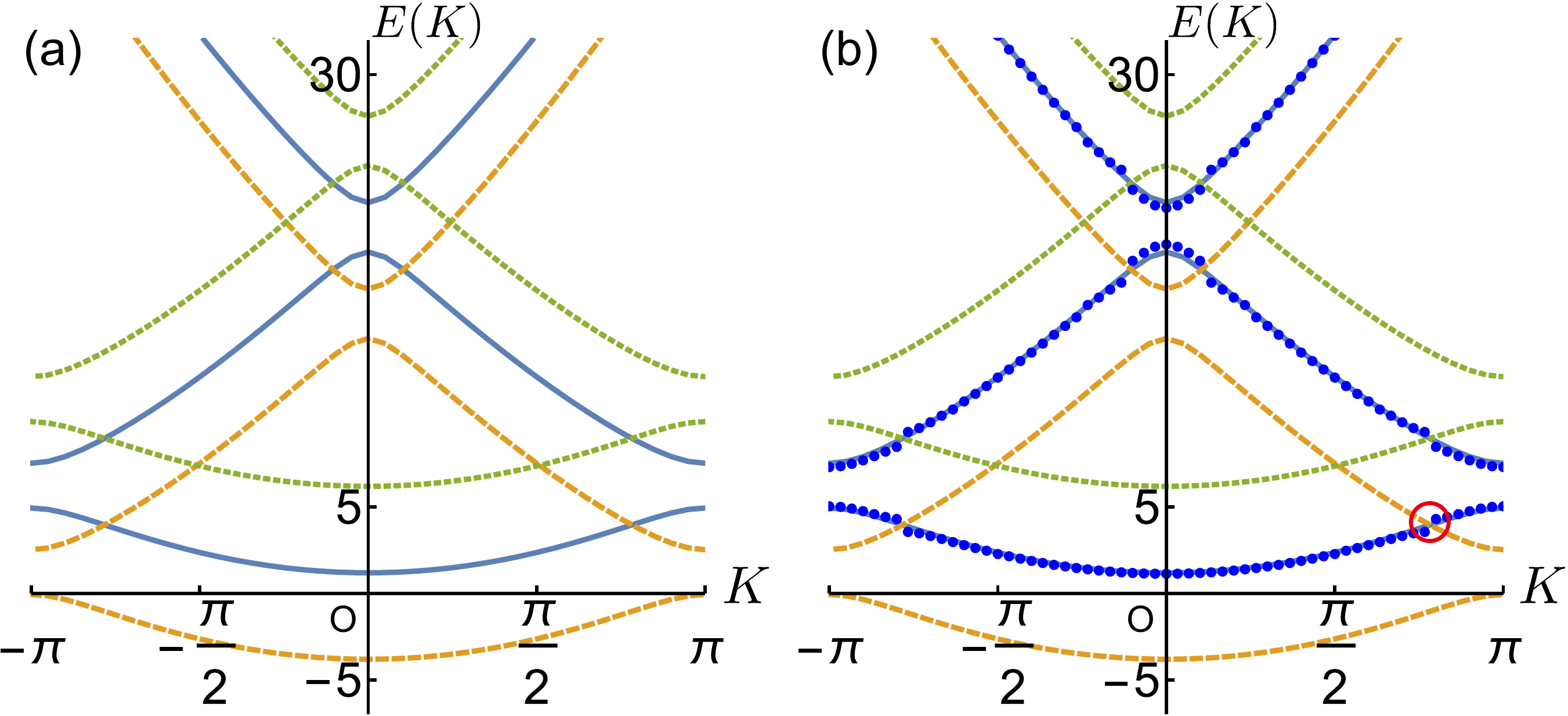}
\caption{(Color Online)
(a) Quasienergy dispersion $E(K)$ for the lattice potential strength $P=3$ in the absence of the coupling between
the electric field and the electron, $\alpha_0=0$.
Each line belongs to the Floquet band with $N=0$ (solid), $-1$ (dashed), and $1$ (dotted),
and $\Omega=5$.
(b) Quasienergy dispersion $E(K)$ for $P=3$, $\alpha_0=0.1$, and $\Omega=5$
(filled circle).
Only the quasienergies corresponding to $N=0$ are plotted.
}
\label{fig:fig2}
\end{figure}
\end{center}

In the absence of the coupling between the electron and the electric filed,
or $\alpha_0=0$,
the quasienergy dispersion
obtained from Eq.~\eqref{eq:kpdisp} is shown in Fig.~\ref{fig:fig2} (a).
Here we also plot the Floquet bands for $N=\pm1$ in addition to the original band $N=0$.
We note again that these states represent the same states as $N=0$
since we work in the extended zone scheme in the energy direction.
Comparing with Fig.~\ref{fig:fig1} (a),
we notice that the periodic potential gives rise to the energy gaps openning at $K=0,\pm \pi$
where two energy bands touch each other
in the absence of the periodic potential.

The coupling between the electron and the electric field
together with the periodic potential
gives rise to the anticrossings of the quasienergy dispersion
at the crossing points of the different Floquet bands.
Figure~\ref{fig:fig2} (b) shows the quasienergy $E(K)$
for $N=0$, $\alpha_0=0.1$, $\Omega=5$, and $P=3$
obtained from Eq.~\eqref{eq:secular2} with the restriction of $N=0,\pm1,\dots,\pm5$.
The quasienergy $E(K)$ is close to the one for $\alpha_0=0$ with $N=0$
except for the vicinity of the crossing points with $N=\pm1$.
One can also find small jumps of the data at the crossing points between
$N=0$ and $N=\pm2$, but cannot between $N=0$ and $N=\pm3,\dots,\pm5$.
This is because the coupling between the Floquet bands is proportional to $\besdiff^{(\pm)}_{N'N}$
(see Eq.~\eqref{eq:defWmat}),
which rapidly decreases as $|N'-N|$ increases.

\begin{center}
\begin{figure}
\includegraphics[width=7cm]{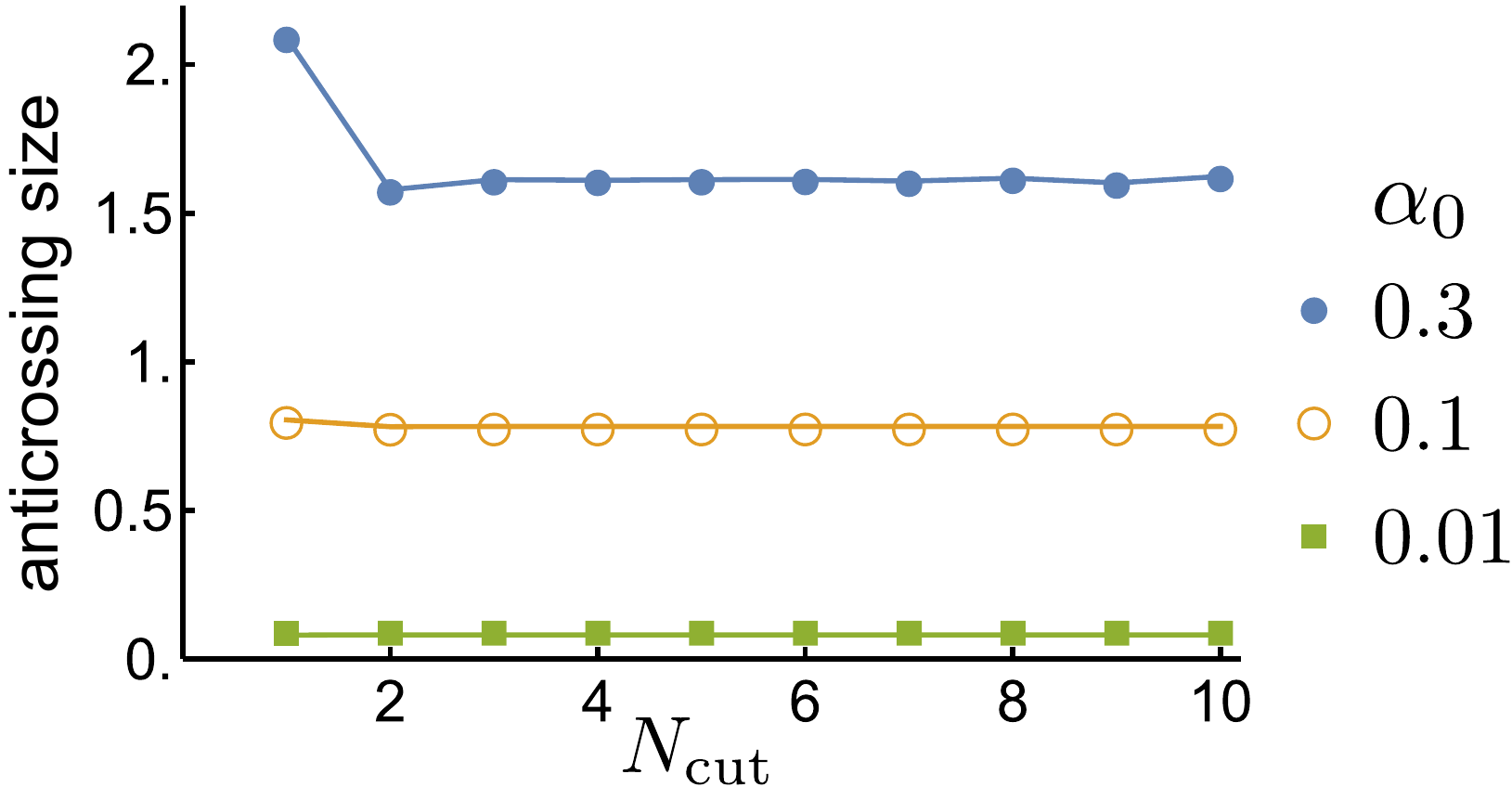}
\caption{
(Color Online)
Anticrossing size at the point indicated by a circle in Fig.~\ref{fig:fig2}(b)
plotted against the cutoff $N_\text{cut}$ ranging from 1 to 10
calculated for $P=3$ and $\Omega=5$.
Each set of the data corresponds to the different coupling strength
between the electron and the electric field:
$\alpha_0=0.3$ (filled circle), 0.1 (open circle), and 0.01 (filled square).
}
\label{fig:fig3}
\end{figure}
\end{center}

The cutoff for $N$ in solving Eq.~\eqref{eq:secular2}
to obtain the $N=0$ band
is justified for a moderate value of $\alpha_0$.
To confirm this, we investigate the quasienergy
by solving Eq.~\eqref{eq:secular2} for $\Omega=5$ and $P=3$
with $N=0,\pm1,\pm2,\dots,\pm N_\text{cut}$, where $N_\text{cut}$ is a varying cutoff. 
Figure~\ref{fig:fig3} shows the size of the anticrossing
at the point indicated by a cirlce in Fig.~\ref{fig:fig2}(b)
calculated for $N_\text{cut}$ ranging from 1 to 10.
For the smaller values of $\alpha_0=0.01$ and $0.1$,
$N_\text{cut}=2$ is large enough to achieve a $10^{-4}$ accuracy for the eigenenergy.
For a larger value of $\alpha_0=0.3$,
the quasienergies show oscillations up to $10\%$ magnitude at $N_\text{cut}\le 3$,
and then converge at $N_\text{cut}=5$ within an accuracy of $10^{-3}$. 
We note that $\alpha_0=0.3$ roughly corresponds to $|A_0|=\pi/2$ in the unit of $|e|=1$
since we set $\Omega=5$ in our calculation.

We remark that
it becomes more complicated to
interpret each quasienergy
for $\alpha_0\gtrsim 0.3$
because the coupling to the higher Floquet bands
are not negligible and more anticrossings appear
at many $K$.
In addition, some anticrossings become so large that
it is not simple to keep track of the one-to-one correspondence between
the quasienergy for $\alpha_0\neq0$ and $0$.
Of course, this is a matter of interpretation,
and Eq.~\eqref{eq:secular2} still provides us with the quasienergy dispersion $E(K)$
even for such a large value of $\alpha_0$.


\subsection{High-harmonic components of the electric current}\label{sec:hhc}
Finally we discuss the electric current
\begin{align}\label{eq:current}
j(x,t) = e \left\{ \text{Re}[ \Psi^*(x,t)(-\ii \partial_x) \Psi(x,t)] -e A(t) |\Psi(x,t)|^2 \right\}
\end{align}
carried by the eigenstate $\Psi(x,t)$ [see Eq.~\eqref{eq:floquetexp}],
which works as the source of HHG.
We show below that the electric current involves high-harmonic components
in the presence of the lattice potential.

In the absence of the lattice potential,
the electric current~\eqref{eq:current} does not involve any high harmonics.
This is because the momentum is a good quantum number and
$-\ii \partial_x \Psi(x,t) = k \Psi(x,t)$ is satisfied for the eigenstate~\eqref{eq:sol0}.
From this relation, we obtain the electric current as
$j(x,t) = e[k-eA(t)]$,
which is homogeneous and contains only frequencies $\pm\Omega$.

In the presence of the lattice potential,
the Floquet-Bloch eigenstate~\eqref{eq:generalsol}
consists of various momenta
and high harmonics are involved in the current.
To show this,
we first compactify Eq.~\eqref{eq:generalsol} as
\begin{align}\label{eq:generalsol2}
\psi_n(x) = \sum_{N,\sigma=\pm} \mathcal{A}_N^\sigma \varphi_n^{N,\sigma k_N}(x),
\end{align}
where we have introduced the notations $\mathcal{A}_N^+=A_N$ and $\mathcal{A}_N^-=B_N$.
Together with Eq.~\eqref{eq:floquetexp},
we obtain 
the nontrivial paramagnetic part of the electric current~\eqref{eq:current} averaged over the unit cell as
(we have been setting $a=1$ in this section)
\begin{align}
\current(t)
\equiv e\int_0^1 \dd x\text{Re}[\Psi^*(x,t)(-\ii \partial_x) \Psi(x,t)]
= \sum_n \current_n\ee^{\ii n\Omega t}\label{eq:hhc}
\end{align}
with
\begin{align}
\current_n\equiv e\sum_{\substack{N,N' \\ \sigma,\sigma',n}} \sigma k_N W_{\sigma'\sigma}^{N'N}(n) (\mathcal{A}_{N'}^{\sigma'})^* \mathcal{A}_N^\sigma,
\end{align}
where $W_{\sigma'\sigma}^{N'N}(n)$ represents
the following wave-function overlap 
\begin{align}\label{eq:overlaps}
W_{\sigma'\sigma}^{N'N}(n)
&\equiv
\sum_{m}\int_0^1 \dd x
\varphi_{m}^{N',\sigma'k_{N'}}(x)^* \varphi_{m+n}^{N,\sigma k_N}(x).
\end{align}
We note $\current_{-n}=\current_n^*$ since $\current(t)$ is real.
Equation~\eqref{eq:hhc} shows that
the current involves an oscillating component with frequency $n\Omega$ if $\current_n\neq 0$.

We make a brief remark on the symmetry of the high-harmonic current.
First, the inversion symmetry of our problem relates the high-harmonic currents
for the degenerate Floquet-Bloch eigenstates that belong to $\pm K$.
If we introduce the notations $\current_n(\pm K)$ to distinguish these two,
one can easily show
\begin{align}\label{eq:hhc_symm}
\current_n(-K) = (-1)^{n+1} \current_n(K).
\end{align}
Thus the high-harmonic currents for an even $n$ vanish
when summed over $\pm K$.
Second, $\current_n$ is real
since our Hamiltonian~\eqref{eq:ham_org}
is symmetric under the product of the inversion and the time reversal.

The result~\eqref{eq:hhc} reduces to the case of no lattice potential
if we put, for example, $A_N=\delta_{N,0}$ and $B_N=0$, which
correspond to the positive momentum state
with momentum $k_0$.
In this case, one can easily see that $\current_n=+k_0 W^{00}_{++}(n)=k_0 \delta_{n0}$.
Similarly, if we put  $A_N=0$ and $B_N=\delta_{N,0}$, which correspond to
the negative momentum state
with momentum $-k_0$, we obtain $p_n=-k_0W^{00}_{--}(n)=-k_0 \delta_{n0}$.
In both cases, the current does not involve any high harmonics as we have shown above.

The high harmonics are induced
once neither $A_N$ nor $B_N$ vanishes
as caused by the Kronig-Penney potential.
To see this fact, we focus on the contributions to $\current_n$ from $N=N'$ with real $k_N$,
which consist of the diagonal $(\sigma=\sigma')$ and the off-diagonal $(\sigma\neq\sigma')$ parts.
The diagonal parts are proportional to 
$|A_N|^2$ or $|B_N|^2$ and again do not contain high harmonics
since they are proportional to $\delta_{n0}$.
On the other hand, 
the off-diagonal parts are proportional to $A_N^*B_N$ and $B_N^*A_N$
and involve a nontrivial dependence on $n$ given by $J_n(2\alpha_0 k_N)$,
which becomes nonzero for $n>1$ in general.

We note that this type of high harmonics are closely related to 
the high-harmonic distribution~\eqref{eq:dist0def}.
If the high-harmonic distribution $|\varphi_n^{N,k}(x)|^2$ were trivially $\delta_{nN}$,
the wave-function overlap $W^{NN}_{\sigma' \sigma}(n)$ would vanish for any $n>1$.
Thus
the width of the high-harmonic distribution
roughly corresponds to the highest order of harmonics of the electric current.

\begin{center}
\begin{figure}
\includegraphics[width=8cm]{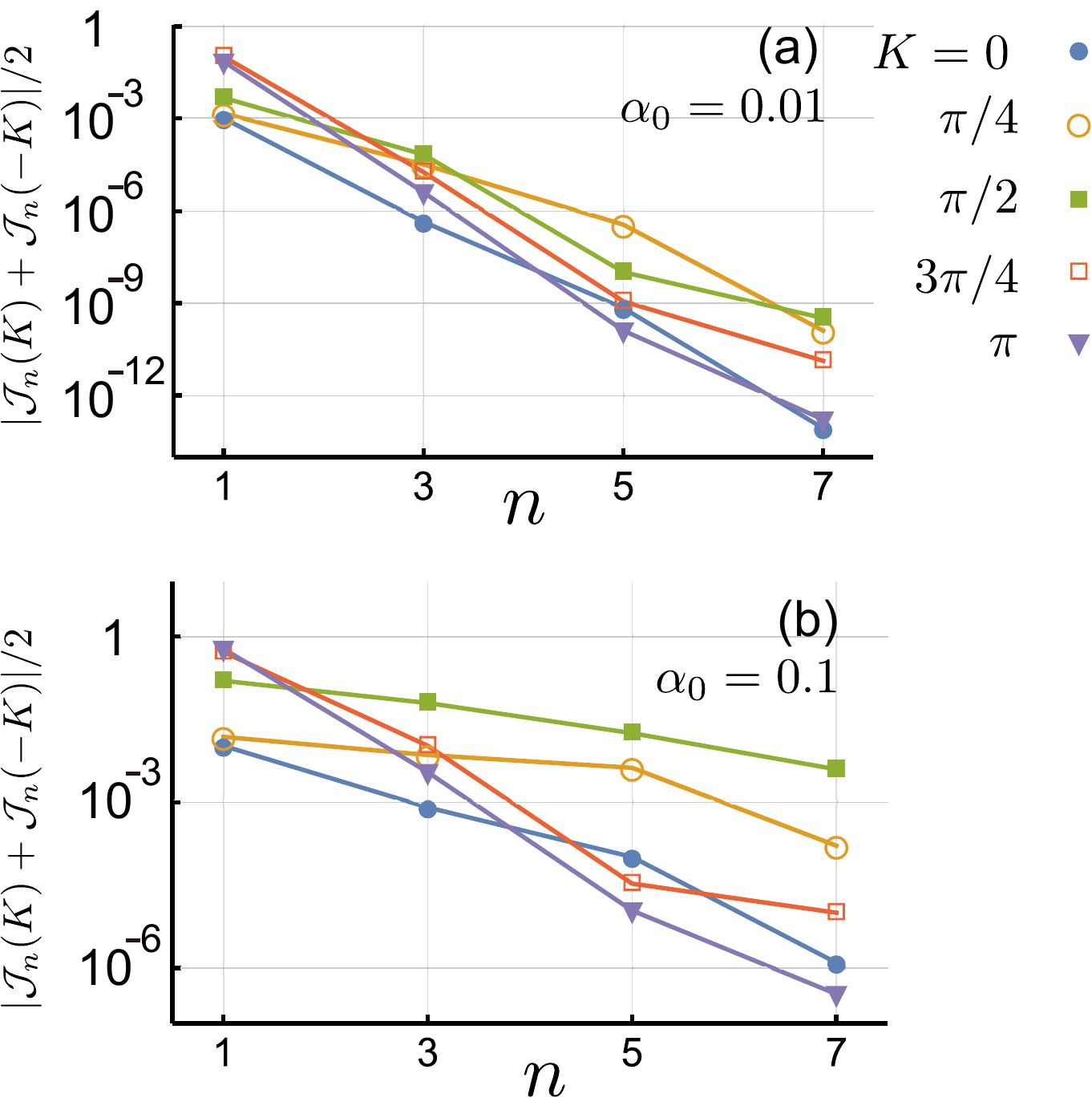}
\caption{
(Color Online)
$K$-resolved high-harmonic spectrum of the electric current,
$|\current_n(K)+\current_n(-K)|/2$ in the unit of $|e|$,
calculated for the Floquet-Bloch eigenstates corresponding to
the lowest band in the absence of the oscillating electric field
(the lowest data points in Fig.~\ref{fig:fig2}(b)).
The two panels are the results for $\alpha_0=0.01$ (a) and 0.1 (b),
and each data set corresponds to the lattice momentum
$K=0$ (filled circle), $\pi/4$ (open circle),
$\pi/2$ (filled square), $3\pi/4$ (open square), and $\pi$ (filled triangle).
}
\label{fig:fig4}
\end{figure}
\end{center}

Actually $\current_n$ also contains the contributions from $N\neq N'$
and numerical calculations are needed to obtain the accurate values.
Figure~\ref{fig:fig4} shows the numerically calculated high-harmonic currents
for the Floquet-Bloch eigenstates, which correspond to the lowest band
in the absence of the oscillating electric field (see the lowest data points in Fig.~\ref{fig:fig2} (b)).
The wave function is renormalized so that $\int_0^1 \dd x |\Psi(x,0)|^2=1$.
Considering Eq.~\eqref{eq:hhc_symm},
we have focused on $|\current_n(K)+\current_n(-K)|/2$ to obtain the $K$-resolved information.
As shown in Fig.~\ref{fig:fig4}, the high-harmonic current at every $K$ decreases exponentially as $n$ increases
for a small coupling $\alpha_0=0.01$ between the electron and the oscillating electric field.
On the other hand, for a larger coupling $\alpha_0=0.1$,
the high-harmonic currents show plateau behaviors around $K=\pi/2$.
Since these plateaus give the largest values for $n=3,5$, and 7,
we also expect a plateau for $\current_n(K)$ summed over $K$, which is the source of HHG.
Thus our result is qualitatively consistent with experiments.

\section{Conclusions}\label{sec:conclusions}
We have revisited the Floquet-Bloch theory
for a one-dimensional electron gas
in the presence of the Kronig-Penney potential in the delta-function limit
and the oscillating electric field.
Taking advantage of the special form of the potential,
we have shown that the Floquet-Bloch eigenstate
is obtained from an eigenvalue problem for a single-infinite matrix,
which is much simpler than the double-infinite matrix needed for generic
periodic potentials.
We have numerically solved the problem to
obtain the quasienergy dispersion $E(K)$,
and shown that
it has the anticrossings at the crossing points of the Floquet bands
as well as the band gaps at the edges and the center  of the Brillouin zone.
We have also confirmed that the quasienergy $E(K)$ is obtained 
at high precision especially for a small amplitude of the vector potential
$|A_0|\lesssim \pi/2$ in the unit of $|e|=1$.
We have then calculated the high-harmonic components
of the electric current for the Floquet-Bloch eigenstates thus obtained,
showing that a plateau appears when the coupling between the electron
and the oscillating electric field is strong enough.

Application to HHG in solids
especially in the nonperturbative regime
is of great interest.
As shown by
the time-dependent Schr\"{o}dinger equation approach~\cite{Ikemachi2017},
the multiple-plateau structure of HHG is related to
the characteristic amplitudes of the vector potential $|A_0|=\pi,2\pi,3\pi,\dots$
in the unit of $|e|=1$.
Once it is achieved to control our calculations at such a large amplitude,
our approach will contribute to the better understand of HHG in solids.
In addressing the strong coupling regime by the Floquet-Bloch theory,
our single-infinite matrix formulation is expected to be of great benefit
compared with the ordinary double-infinite matrix one.

\section*{Acknowledgements}
Fruitful discussions with Hirokazu Tsunetsugu
are gratefully acknowledged.
This work was supported by
JSPS KAKENHI Grant Nos.~JP16H06718 and JP18K13495.


\appendix

\section{Floquet-Green function approach}

In this appendix,
we review the original approach to solve
Eqs.~\eqref{eq:schroe1} and \eqref{eq:schroe2}
on the basis of the Floquet-Green functions
by Faisal and Genieser~\cite{Faisal1989,Faisal1991}.
Unfortunately, their results contain mistakes,
which we point out in the following.

The unperturbed Floquet-Green function $G^0_{nn'}(x,x')$ is defined by
\begin{align}\label{eq:defG0}
(E-H^0)G^0_{nn'}(x,x') = \delta_{nn'}\delta(x-x')
\end{align}
and its explicit form is given by
\begin{align}\label{eq:G0exp}
G_{nn'}^0(x,x')=
\sum_{N}\int_{-\infty}^\infty \frac{\dd k}{2\pi} \frac{\varphi^{Nk}_n(x) \varphi^{Nk}_{n'}(x')^*}{E-(k^2/2+N\Omega)+i0},
\end{align}
where $\varphi^{Nk}_n(x)$ is the eigenstate of $H^0$ (see Eq.~\eqref{eq:sol0}).
In this appendix, the summation is taken over $\mathbb{Z}$ if its range is not specified.
The Floquet-Green function $G^0_{nn'}(x,x')$
enables us to transform Eq.~\eqref{eq:schroe1} as
\begin{align}\label{eq:inverted}
\psi_n(x) = \sum_{n'}\int_{-\infty}^\infty \dd x'\, G_{nn'}^0(x,x')V(x')\psi_{n'}(x').
\end{align}

Once $G^0_{nn'}(x,x')$ is known,
the quasienergy is obtained as follows.
We invoke the Bloch theorem
and assume that our eigenstate is written as
\begin{align}\label{eq:blochphi}
\psi_n(x) = \ee^{\ii Kx} \phi^K_n(x),
\end{align}
where $K$ is a lattice momentum
and $\phi^K_n(x)$ is a periodic function: $\phi_n^K(x+a)=\phi_n^K(x)$.
Similarly to the argument in Sec.~\ref{sec:solution},
this leads to the quasienergy $E$ for each $K$,
which is denoted as $E(K)$.
By substituting Eqs.~\eqref{eq:kppot} and \eqref{eq:blochphi} into Eq.~\eqref{eq:inverted},
we have
\begin{align}\label{eq:faisal0}
\phi_n^K(0) = \frac{P}{2a}\sum_{p,n'}G^0_{nn'}(0,pa)\ee^{\ii pKa}\phi^K_n(0),
\end{align}
where we have used the periodicity $\phi^K_n(pa)=\phi^K_n(0)$ for any $p\in\mathbb{Z}$.
This is a linear homogeneous equation for $\phi^K_n(0)$
and the secular equation determines the quasienergy $E(K)$.

To obtain the simplified expression for $G^0_{nn'}(x,x')$,
Faisal and Genieser~\cite{Faisal1989,Faisal1991} performed the integration over $k$
on the right-hand side of Eq.~\eqref{eq:G0exp}
by invoking residue calculus.
Focusing on the factor $\ee^{\ii k(x-x')}$ in the integrand,
they added a contour integral along the infinitely-large semicircle on the upper- or lower-half
of the complex $k$ plane depending on $x\ge x'$ or $x<x'$, respectively,
assuming that this additional contour integral gives no contribution.
Then they calculated the residues of the integrand at the poles
encircled by the composite contour consisting of the real axis and the semicircle,
obtaining,
for $x\ge x'$,
\begin{align}
&G_{nn'}^\FG(x,x')\notag\\
&=-\ii \sum_{N} \frac{1}{k_N}J_{n-N}(\alpha_0k_N)J_{n'-N}(\alpha_0k_N)\ee^{\ii k_N (x-x')} \label{eq:fg1}
\end{align}
and, for $x<x'$,
\begin{align}
&G_{nn'}^\FG(x,x')\notag\\
&=-\ii \sum_{N} \frac{1}{k_N}J_{n-N}(-\alpha_0k_N)J_{n'-N}(-\alpha_0k_N)\ee^{-\ii k_N (x-x')},\label{eq:fg2}
\end{align}
where $k_N$ is defined in Eq.~\eqref{eq:defkN}
and the notation $G_{nn'}^\FG(x,x')$ is used to be distinguished from the true $G^0_{nn'}(x,x')$.

However,
this line of reasoning contains a mistake
because the contour integral along the infinitely-large semicircle
is not negligible for small $|x-x'|$ due to $J_{n-N}(\alpha_0k)J_{n'-N}(\alpha_0k)$
in the integrand.
This becomes manifest if one represents the Bessel function $J_{n-N}(\alpha_0k)$
in terms of the Hankel functions as $J_{n-N}(\alpha_0k)=[H^{(1)}_{n-N}(\alpha_0k)+H^{(2)}_{n-N}(\alpha_0k)]/2$
and notes their asymptotic forms
\begin{align}
H_{n-N}^{(1,2)}(\alpha_0k) \sim \sqrt{\frac{2}{\pi \alpha_0k}}\exp \left[ \pm \ii \left( \alpha_0k -\frac{\pi(n-N)}{2}-\frac{\pi}{4} \right) \right]
\end{align}
in $|\alpha_0 k|\to \infty$.
Especially for $x=x'$,
neither of 
the contour integrals along the semicircles on the upper- and lower-half complex $k$ plane 
converges due to the coexistence of the terms proportional to $\ee^{\pm2\ii \alpha_0 k}$ in the integrand.

As a result,
Eqs.~\eqref{eq:fg1} and \eqref{eq:fg2} do not satisfy Eq.~\eqref{eq:defG0}
at $x=x'$.
Namely, Eq.~\eqref{eq:defG0} implies
\begin{widetext}
\begin{align}\label{eq:g0connection}
&\frac{1}{2}\partial_xG_{nn'}^0(x'+0,x')-\frac{1}{2}\partial_xG_{nn'}^0(x'-0,x')\notag\\
&-\ii \Omega\frac{\alpha_0}{2}\left\{
\left[G^0_{n+1,n'}(x'+0,x') +G^0_{n-1,n'}(x'+0,x')\right]
-\left[G^0_{n+1,n'}(x'-0,x') +G^0_{n-1,n'}(x'-0,x')\right] \right\}
=\delta_{nn'},
\end{align}
\end{widetext}
which is obtained by integrating both sides of Eq.~\eqref{eq:defG0}
over $x\in [x'-\epsilon,x'+\epsilon]$ and taking the limit of $\epsilon \downarrow0$.
When evaluated for $G^{\FG}_{nn'}(x,x')$ instead of $G^0_{nn'}(x,x')$,
the left-hand side of Eq.~\eqref{eq:g0connection}
turns out to be
\begin{align}
&\frac{1}{2}\sum_N \left(1-\frac{(n-N)\Omega}{k_N^2}\right)\notag\\
&\qquad\times[1+(-1)^{n-n'}] J_{n-N}(\alpha_0k_N)J_{n'-N}(\alpha_0k_N),
\end{align}
which is not equal to $\delta_{nn'}$.
Thus we have shown $G^\FG_{nn'}(x,x')\neq G^0_{nn'}(x,x')$.

We note that the Floquet-Green functions~\eqref{eq:fg1} and \eqref{eq:fg2}
work in the absence of the oscillating electric field.
In this case, we do not have the index $n$,
$H^0$ is just $-\partial_x^2/2$,
and the $G^\FG(x,x')$ reduces to
$-\ii \ee^{\ii k_0 (x-x')}/k_0$ for $x\ge x'$
and $-\ii \ee^{-\ii k_0 (x-x')}/k_0$ for $x<x'$.
As for Eq.~\eqref{eq:g0connection},
we do not have the term proportional to $\Omega$,
and the condition is satisfied for $G^\FG(x,x')$.
One can easily show that
Eq.~\eqref{eq:faisal0} reduces to
\begin{align}
\phi^K(0) = \frac{P}{2k_0 a} \frac{\sin k_0 a}{\cos Ka -\cos k_0a} \phi^K(0),
\end{align}
which gives the dispersion relation~\eqref{eq:kpdisp}
for $\phi^K(0)\neq0$.

At present, the author has not found the proper simplified form of $G^0_{nn'}(x,x')$ yet.

%

\end{document}